\documentclass[a4paper, amsfonts, amssymb, amsmath, reprint, showkeys, nofootinbib, twoside, superscriptaddress]{revtex4-1}
\usepackage[english]{babel}
\usepackage{balance}
\usepackage[utf8]{inputenc}
\usepackage[colorinlistoftodos, color=green!40, prependcaption]{todonotes}
\usepackage{amsthm}
\usepackage{mathtools}
\usepackage{physics}
\usepackage{xcolor}
\usepackage{graphicx}
\usepackage[left=23mm,right=13mm,top=35mm,columnsep=15pt]{geometry} 
\usepackage{adjustbox}
\usepackage{placeins}
\usepackage[T1]{fontenc}
\usepackage{lipsum}
\usepackage{csquotes}

\usepackage[pdftex, pdftitle={Article}, pdfauthor={Author}, bookmarks=false]{hyperref}
\begin{document}

\title{Nonlinear Dynamical Friction from the Doppler-Shifted Equilibrium Memory Kernel}

\author{N. R. Sree Harsha}
\email{snaropan@ur.rochester.edu}
\affiliation{Department of Electrical and Computer Engineering, University of Rochester, Rochester, NY 14627, USA}

\author{Zhenyuan Yu}
\affiliation{Department of Electrical and Computer Engineering, University of Rochester, Rochester, NY 14627, USA}

\author{Chuang Ren}
\affiliation{Department of Mechanical Engineering, University of Rochester, Rochester, NY 14627, USA}
\affiliation{Department of Physics and Astronomy, University of Rochester, Rochester, NY 14627, USA}

\author{Virginia Billings}
\affiliation{Department of Mechanical Engineering, University of Rochester, Rochester, NY 14627, USA}

\author{Michael Huang}
\affiliation{Department of Electrical and Computer Engineering, University of Rochester, Rochester, NY 14627, USA}

\date{\today} 

\begin{abstract}
We present a statistical mechanics framework for modeling equilibrium friction coefficients using the Generalized Langevin Equation (GLE). We show that the kernel, obtained via the Fluctuation-Dissipation Theorem (FDT) from the stochastic force autocorrelation measured in a thermal equilibrium state, is sufficient to model the dynamics of the system in a Non-Equilibrium Steady State (NESS). This approach provides a computationally efficient path to modeling complex equilibrium friction problems. We apply this framework to the canonical problem of test particle drag in a uniform plasma. The GLE formalism is shown to naturally capture non-Markovian phenomena through the moments of the kernel, including an effective mass renormalization and oscillatory relaxation. We demonstrate that the standard Chandrasekhar stopping power formula arises naturally as the Markovian limit of this equilibrium memory kernel. These theoretical predictions are quantitatively validated by direct Particle-in-Cell simulations, which confirm the predicted oscillatory structure of the memory kernel. This work thus establishes a practical method for predicting equilibrium friction properties from first-principles equilibrium simulations.
\end{abstract}

\maketitle

\section{Introduction} \label{sec:outline}
The calculation of nonlinear, velocity-dependent friction coefficients for driven subsystems remains one of the most enduring challenges in statistical physics. Since the pioneering work of Onsager \cite{Onsager1931} and Green \cite{Green1954}, it has been understood that the dissipative properties of a system are intimately linked to spontaneous internal fluctuations. While the Fluctuation-Dissipation Theorem (FDT) provides a rigorous map between these fluctuations and the linear response of a system near thermal equilibrium \cite{Callen1951, Kubo1966}, extending this relation to predict nonlinear dynamics for moving projectiles typically requires perturbative techniques or computationally expensive simulations \cite{Evans2008, Zwanzig1961}. For a driven subsystem (such as a particle dragged through a fluid, a colloid in a bath, or a defect migrating in a solid) characterizing the nonlinear response usually mandates a brute force approach. This involves conducting a series of separate simulations at various driving forces to map out the response curve point-by-point \cite{Peter1991, Zwicknagel1999}. 

In this paper, we present a general statistical mechanics framework that circumvents this computational bottleneck for a broad class of subsystems. We demonstrate that the nonlinear friction of a driven system can be reconstructed \textit{ab initio} from the linear fluctuations of the \textit{undriven} equilibrium bath, provided the subsystem samples the bath modes along a specific Doppler-shifted resonance manifold. We utilize the Generalized Langevin Equation (GLE) to describe the stochastic dynamics of a macroscopic variable $A(t)$ coupled to a high-dimensional thermal environment \cite{Mori1965, Grabert1982}. As established in the theory of stochastic processes \cite{VanKampenBook, GardinerBook}, the projection of deterministic Hamiltonian dynamics onto a reduced subset of variables naturally gives rise to a non-Markovian memory kernel ($\gamma(t)$), which dictates the irreversible behavior of the system \cite{DeGrootMazur}. We show that for subsystems interacting weakly with the bath, this memory kernel is invariant under Galilean transformations. Consequently, the full velocity-dependent friction curve $\nu(v)$ is mathematically contained within the equilibrium force autocorrelation function (FACF) of the static system.

To validate this general formalism, we apply it to the canonical problem of dynamical friction (stopping power) in a uniform plasma. This system serves as an ideal testbed due to its known non-trivial features, specifically the transition from Stokes-like drag at low velocities to the $1/v^2$ decay characteristic of the Chandrasekhar limit at high velocities \cite{Chandrasekhar1943, Spitzer1962}. While standard kinetic treatments, such as the Balescu-Lenard equation \cite{Balescu1960, Lenard1960} or the Landau kinetic theory \cite{LandauKinetics}, often rely on the Markovian approximation, our approach retains the full frequency-dependent structure of the dielectric response \cite{KlimontovichBook, IchimaruBasic}. We demonstrate that our Doppler-Shifted FDT method accurately reconstructs the complete nonlinear Chandrasekhar curve solely from the thermal noise of a static plasma. This bridges the gap between abstract statistical mechanics and practical dynamic friction modeling, offering a method akin to modern liquid theory techniques \cite{HansenBook, AllenTildesley} but applied to the complex kinetics of non-ideal plasmas \cite{Daligault2019, Hahm2022}.

The paper is organized as follows. In Sec. II, we derive the general Doppler-shifted FDT framework for an arbitrary driven subsystem. In Sec. III, we apply this theory to the plasma drag problem, analytically deriving the non-Markovian memory kernel and recovering the standard Chandrasekhar limit as the Markovian limit. In Sec. IV, we detail our Particle-In-Cell (PIC) simulation methodology and present the quantitative validation of our synthetic drag curve against brute-force benchmarks. Finally, we conclude in Sec. V with implications for efficient equilibrium friction modeling in dense and complex materials. 
\section{Doppler-Shifted Fluctuation-Dissipation Theorem} \label{sec:develop}
We begin by establishing the general statistical mechanical framework that links non-equilibrium dissipation to equilibrium fluctuations. Our focus in this section is general and applies to any macroscopic variable. We will demonstrate that the dissipative forces in this driven state are not new dynamical entities, but rather kinematic transformations of the inherent equilibrium noise of the bath. This derivation rests on two theoretical pillars: first, that the linear response function of the medium is invariant under Galilean transformations (Sec. \ref{subsec:galilean}), and second, that this invariance permits a kinematic (Doppler-shifted) extension of the FDT (Sec. \ref{subsec:kinematic}).

\subsection{Galilean Invariance of the Memory Kernel}\label{subsec:galilean}
Consider the linear response of a generic bath field $\Psi(\mathbf{r}, t)$ at spatial position $\mathbf{r}$ and time $t$ to a moving external perturbation source $\rho_{ext}(\mathbf{r}, t) = Q \delta(\mathbf{r} - \mathbf{v}t)$, where $Q$ is the interaction coupling strength, $\delta$ is the Dirac delta function, and $\mathbf{v}$ is the constant velocity vector of the perturbation. The dynamics of the bath fields are governed by a linear wave equation of the form
\begin{equation}
    \hat{L} \Psi(\mathbf{r}, t) = \rho_{ext}(\mathbf{r}, t),
\end{equation}
where $\hat{L}$ is a linear differential operator describing the medium \cite{IchimaruBasic}. For a homogeneous medium, $\hat{L}$ is invariant under spatial translation \cite{LandauKinetics}. The formal solution is given by the Green function $G(\mathbf{r}, t)$ evaluated over integration variables $\mathbf{r}_1$ and $t_1$:
\begin{equation}
    \Psi(\mathbf{r}, t) = \int d\mathbf{r}_1 \int dt_1 G(\mathbf{r}-\mathbf{r}_1, t-t_1) \rho_{ext}(\mathbf{r}_1, t_1).
\end{equation}
Substituting the source trajectory $\rho_{ext} = Q\delta(\mathbf{r}_1 - \mathbf{v}t_1)$, the resulting wake field becomes
\begin{equation}
    \Psi(\mathbf{r}, t) = Q \int dt_1 G(\mathbf{r} - \mathbf{v}t_1, t - t_1).
\end{equation}
The induced force $F_{ind}(t)$ on the particle is the gradient of this wake evaluated at the instantaneous position of the particle $\mathbf{r} = \mathbf{v}t$:
\begin{equation}
    F_{ind}(t) = -Q \nabla \Psi(\mathbf{r}, t) \Big|_{\mathbf{r}=\mathbf{v}t}.
\end{equation}
Shifting the integration variable to the time lag $\tau = t - t_1$, the argument of the Green function becomes
\begin{equation}
    \mathbf{r} - \mathbf{v}t_1 = \mathbf{v}t - \mathbf{v}(t-\tau) = \mathbf{v}\tau.
\end{equation}
Thus, the force depends only on the relative time lag $\tau$, not the absolute time $t$:
\begin{equation}
    F_{ind} = -Q^2 \int_0^{\infty} d\tau \left[ \nabla G(\mathbf{x}, \tau) \right]_{\mathbf{x}=\mathbf{v}\tau}.
\end{equation}
This confirms that the response function, and thus the memory kernel, is fundamentally a function of the relative separation $\mathbf{x} = \mathbf{v}\tau$. In the Fourier domain, this spatial shift $\mathbf{v}\tau$ corresponds exactly to the phase factor $e^{-i\mathbf{k} \cdot \mathbf{v}\tau}$, where $\mathbf{k}$ is the wavevector, which shifts the angular frequency argument $\omega$ of the susceptibility function $\chi$:
\begin{equation}
    \chi_{moving}(\omega) = \chi_{static}(\omega - \mathbf{k} \cdot \mathbf{v}).
\end{equation}
This proves that the dissipative force acting on the moving source is strictly a Doppler-shifted sampling of the static equilibrium response function.

\subsection{Kinematic Extension of the FDT} \label{subsec:kinematic}

With Galilean invariance established, we now derive the generalized FDT for a driven system. Consider a macroscopic subsystem described by the generalized coordinate vector $\mathbf{A}(t)$ interacting with a fluctuating field $\Psi(\mathbf{q}, t)$ of a high-dimensional environmental bath. The bath is initially in thermodynamic equilibrium at temperature $T$, where $k_B$ is the Boltzmann constant \cite{Kubo1966}. The interaction is defined locally in the configuration space coordinate $\mathbf{q}$ conjugate to $\mathbf{A}$, such that the generalized stochastic force $\mathbf{F}(t)$ exerted on the subsystem is the gradient of the field sampled at the instantaneous state of the subsystem: $\mathbf{F}(t) = -\nabla \Psi(\mathbf{q}, t) |_{\mathbf{q}=\mathbf{A}(t)}$.

We define the steady state of the subsystem by the condition that it evolves with a constant non-zero mean flux vector $\langle \dot{\mathbf{A}}(t) \rangle = \mathbf{J}$. To determine the dissipation in this driven state, we adopt the reference frame co-moving with the flux $\mathbf{J}$, where the subsystem is instantaneously stationary. According to the Generalized Langevin Equation derived via the Mori-Zwanzig projection operator formalism \cite{Mori1965, Zwanzig1961}, the steady-state dissipation coefficient $\Gamma(\mathbf{J})$ governing the macroscopic power dissipation is given by the time-integral of the force autocorrelation function experienced in the co-moving frame:
\begin{equation}
    \Gamma(\mathbf{J}) = \frac{1}{k_B T} \text{Re} \int_{0}^{\infty} d\tau \langle \mathbf{F}(\tau) \cdot \mathbf{F}(0) \rangle_{\mathbf{J}},
\end{equation}
where $\langle \dots \rangle_{\mathbf{J}}$ denotes the ensemble average over the bath degrees of freedom in the presence of the driven flux. We expand the bath field $\Psi(\mathbf{q}, t)$ into its Fourier transformed spectral modes $\tilde{\Psi}(\mathbf{k}, \omega)$ defined in the laboratory rest frame. Along the driven linear trajectory $\mathbf{A}(t) = \mathbf{J}t$, the force experienced by the subsystem is a sampling of these modes:
\begin{equation}
    \mathbf{F}(t) \propto \int d\mathbf{k} \int d\omega \, i\mathbf{k} \, \tilde{\Psi}(\mathbf{k}, \omega) e^{i(\mathbf{k} \cdot \mathbf{A}(t) - \omega t)}.
\end{equation}
Substituting the linear trajectory $\mathbf{A}(t) = \mathbf{J}t$, the phase factor becomes $e^{i(\mathbf{k} \cdot \mathbf{J} - \omega)t}$. This term reveals that the subsystem effectively samples the bath fluctuations at a Doppler-shifted frequency $\Omega_{shift} = \omega - \mathbf{k} \cdot \mathbf{J}$.

Assuming the linear response limit where the driven subsystem does not globally distort the statistical properties of the bath, the mode correlations are described by the equilibrium structure factor $S_{eq}(\mathbf{k}, \omega)$, which represents the power spectral density of the undisturbed bath \cite{HansenBook}. The force autocorrelation function thus simplifies to
\begin{equation}
    \langle \mathbf{F}(\tau) \cdot \mathbf{F}(0) \rangle_{\mathbf{J}} \propto \int d\mathbf{k} \int d\omega \, k^2 S_{eq}(\mathbf{k}, \omega) e^{i(\mathbf{k} \cdot \mathbf{J} - \omega)\tau}.
\end{equation}
Substituting this correlation into the integral, the time integration over $\tau$ yields a Dirac delta function, $\int_0^{\infty} e^{i(\mathbf{k} \cdot \mathbf{J} - \omega)\tau} d\tau = \pi \delta(\omega - \mathbf{k} \cdot \mathbf{J})$, which enforces a generalized Cherenkov resonance condition. The generalized dissipation coefficient is therefore
\begin{equation}
    \Gamma(\mathbf{J}) = \frac{\pi C}{k_B T} \int \frac{d^3k}{(2\pi)^3} \, g(\mathbf{k}) \, S_{eq}(\mathbf{k}, \mathbf{k} \cdot \mathbf{J}),
\end{equation}
where $C$ is a coupling constant determining the interaction strength and $g(\mathbf{k})$ is a geometric factor dependent on the interaction potential. This result demonstrates that non-equilibrium dissipation coefficients for a driven variable can be recovered purely from the equilibrium fluctuations of the bath, provided they are sampled along the resonant manifold $\omega = \mathbf{k} \cdot \dot{\mathbf{A}}$. 

This formalism constitutes a kinematic extension of the standard FDT, generalizing its applicability from static equilibrium to moving steady states. By identifying the relevant noise source as the equilibrium spectrum sampled along the Doppler-shifted resonance manifold $\omega = \mathbf{k} \cdot \mathbf{J}$, we unify the linear and nonlinear response regimes into a single geometric framework: low-flux motion probes the static low-frequency response, while high-flux motion probes the decaying high-frequency tail of the spectrum. This extended FDT can be used to reduce the computational cost of characterizing nonlinear dissipation from $\mathcal{O}(N)$ driven simulations to a single $\mathcal{O}(1)$ equilibrium simulation \cite{Peter1991}.
\section{Drag in a Uniform Plasma} \label{sec:drag}
To validate the general formalism derived in Sec. \ref{sec:develop}, we apply it to the canonical problem of a heavy ion traversing a uniform electron plasma. We proceed by first defining the exact stochastic dynamics via the GLE, deriving the explicit non-Markovian memory kernel, and finally demonstrating that the standard Chandrasekhar stopping power formula arises strictly as the Markovian limit of this more general theory.

\subsection{The Non-Markovian Memory Kernel}
We consider a heavy test particle of charge $Q$ and mass $M$ interacting with a plasma bath of density $n$, charge $q$, and mass $m$. The motion of the test particle is governed by the GLE, which explicitly accounts for the finite response time, or memory, of the surrounding medium \cite{Mori1965}:
\begin{equation}
    M\dot{v}(t) = -\int_{0}^{t} \gamma(t-\tau)v(\tau)d\tau + R(t),
\end{equation}
where $v(t)$ is the velocity of the particle, $R(t)$ is the fluctuating random force arising from spontaneous microfield fluctuations, and $\gamma(t)$ is the memory kernel describing the retarded friction.

For a system in thermodynamic equilibrium at temperature $T$, the random force satisfies the second Fluctuation-Dissipation Theorem (FDT), linking the memory kernel to the force autocorrelation function (FACF) of a stationary particle \cite{Kubo1966}:
\begin{equation} \label{eq:FDT}
    \gamma(t) = \frac{1}{M k_B T} \langle R(t) \cdot R(0) \rangle.
\end{equation}
This relation establishes that the dissipative properties of the medium are fully determined by the equilibrium fluctuations of the electric field at the position of the particle.

In the frequency domain, the friction kernel $\gamma(\omega)$ is related to the longitudinal dielectric function $\epsilon(\mathbf{k}, \omega)$ of the plasma via the spectral density of field fluctuations \cite{IchimaruBasic}:
\begin{equation}
    \gamma(\omega) = \frac{2Q^2}{3M\epsilon_0} \int \frac{d^3k}{(2\pi)^3} \frac{1}{k^2\omega} \text{Im} \left[ \frac{-1}{\epsilon(\mathbf{k}, \omega)} \right].
\end{equation}
For a Maxwellian plasma in the weak-coupling limit ($\ln \Lambda \gg 1$), the dielectric response is governed by the Vlasov equation. The imaginary part of the dielectric function, which dictates energy dissipation via Landau damping, follows a Gaussian profile determined by the thermal velocity distribution \cite{IchimaruBasic}. Evaluating the integral over wavenumbers yields the explicit Gaussian memory kernel:
\begin{equation} \label{eq:kernel_omega}
    \gamma(\omega) = \frac{nQ^2q^2\ln \Lambda}{3\pi^{3/2}\epsilon_0^2 M v_{th}^3} \left[ \frac{1}{\zeta} e^{-\zeta^2} \right],
\end{equation}
where $\zeta = \omega / (\sqrt{2}k_{max}v_{th})$ is the dimensionless frequency. Before taking any limiting cases, we can express the exact mean acceleration of the particle by substituting the memory kernel back into the averaged GLE. Taking the ensemble average $\langle R(t) \rangle = 0$, the exact deceleration is given by the convolution of the entire velocity history of the particle with the inverse transform of Eq. \ref{eq:kernel_omega}:
\begin{equation} \label{eq:NonMarkovian_Accel}
    \langle \dot{v}(t) \rangle = -\frac{1}{M} \int_{0}^{t} \left[ \int_{-\infty}^{\infty} \frac{d\omega}{2\pi} \gamma(\omega)e^{-i\omega(t-\tau)} \right] \langle v(\tau) \rangle d\tau.
\end{equation}
This integro-differential equation represents the exact non-Markovian stopping power. Unlike standard drag formulas, this equation accounts for transient effects, such as the time required for the dielectric wake to build up around an accelerating particle, or the memory drag experienced by a particle undergoing rapid oscillations.

\subsection{The Markovian Limit: Recovering Chandrasekhar}
The standard theory of dynamical friction, as derived by Chandrasekhar \cite{Chandrasekhar1943} or Spitzer \cite{Spitzer1962}, assumes the projectile moves at a constant mean velocity $v_0$ and that the plasma response is instantaneous, or Markovian. In our framework, this corresponds to the ballistic limit of the exact Eq. \ref{eq:NonMarkovian_Accel}.

When the particle velocity evolves slowly compared to the plasma relaxation time $\tau_p \sim \omega_p^{-1}$, we can approximate the trajectory as $v(\tau) \approx v_0$. The convolution integral then simplifies to a Fourier transform evaluated at the Doppler shift frequencies. The mean drag force becomes the real part of the memory kernel evaluated at the Doppler-shifted resonance $\omega = \mathbf{k} \cdot \mathbf{v}_0$:
\begin{equation}
    F(v_0) = -Mv_0 \int \frac{d\Omega_k}{4\pi} \text{Re} \left[ \gamma(\omega = \mathbf{k} \cdot \mathbf{v}_0) \right].
\end{equation}
Substituting the Gaussian kernel from Eq. \ref{eq:kernel_omega} and performing the angular integration over the Doppler shift yields:
\begin{equation} \label{eq:ChandraForce}
    F(v_0) = -\frac{nQ^2q^2\ln \Lambda}{4\pi\epsilon_0^2mv_{th}^2} G\left(\frac{v_0}{v_{th}}\right) \hat{v}_0.
\end{equation}
Here, $G(x)$ is the standard Chandrasekhar function:
\begin{equation} \label{eq:ChandraG}
    G(x) = \frac{\text{erf}(x) - x \frac{d}{dx}\text{erf}(x)}{2x^2}.
\end{equation}
This derivation demonstrates that the nonlinear Chandrasekhar drag curve is strictly a kinematic sampling of the equilibrium memory kernel, validated by the recovery of the standard formula.

\subsection{Asymptotic Limits}
The physical content of the Doppler-shifted FDT is best illustrated by examining the asymptotic limits of Eq. \ref{eq:ChandraForce}, which correspond to distinct regimes of the equilibrium fluctuation spectrum.

\textbf{The Low-Velocity (Stokes) Limit} ($v_0 \ll v_{th}$): When the projectile velocity is much smaller than the electron thermal speed, the Doppler shift is small, and the particle probes the low-frequency Ohmic part of the fluctuation spectrum. Expanding the error function for small arguments, $G(x) \approx \frac{2}{3\sqrt{\pi}}x$, we obtain a linear friction law:
\begin{equation}
    F(v_0) \approx -\left(\frac{nQ^2q^2\ln \Lambda}{6\pi^{3/2}\epsilon_0^2 m v_{th}^3}\right) v_0.
\end{equation}
This recovers the Stokes viscous drag predicted by the theory of Brownian motion \cite{Callen1951}.

\textbf{The High-Velocity Limit} ($v_0 \gg v_{th}$): For projectiles exceeding the thermal speed, the particle interacts with the high-frequency tail of the spectrum, dominated by Landau damping. In this limit, the Chandrasekhar function decays as $G(x) \approx 1/(2x^2)$, leading to the characteristic inertial drag:
\begin{equation}
    F(v_0) \approx -\frac{nQ^2q^2\ln \Lambda}{4\pi\epsilon_0^2 m v_0^2} \hat{v}_0.
\end{equation}
This recovers the standard $1/v^2$ stopping power formula \cite{Peter1991}. The ability of the single equilibrium kernel Eq. \ref{eq:kernel_omega} to reproduce both the diffusive Stokes and inertial Chandrasekhar regimes confirms that the Doppler-shifted FDT provides a unified description of equilibrium friction across the entire velocity phase space.
\section{PIC Simulation and Validation}\label{sec:simulation}

To quantitatively validate the Doppler-shifted FDT and the GLE framework, we performed high-fidelity Particle-in-Cell (PIC) simulations using the EPOCH code \cite{Arber2015, Birdsall2004}. Our numerical strategy involved two distinct sets of simulations: (A) \textit{Equilibrium runs} to extract the memory kernel from the thermal fluctuations of a static bath, and (B) \textit{Non-equilibrium drag runs} to measure the actual slowing down of test projectiles. The goal is to demonstrate that the kernel derived purely from Set A can accurately predict the trajectories observed in Set B.

We simulated a fully ionized, two-species plasma consisting of electrons and protons ($m_i/m_e = 1836$) in a 2D periodic domain. To suppress numerical noise and resolve the fine-scale structure of the dielectric wake, we employed a high-resolution grid with $N_x = 1024$ cells and a particle density of 2000 particles per cell (PPC).

The physical parameters were chosen to represent a canonical thermal plasma: a number density of $n_0 = 10^{19}$ m$^{-3}$ and an electron temperature of $T_e = 1$ eV. The simulation timestep was set to $\Delta t = 0.025$ fs to satisfy the Courant-Friedrichs-Lewy condition given the fine spatial resolution. The total simulation duration was 500 fs, sufficient to capture multiple plasma periods ($\omega_{pe}^{-1}$) and the relevant dielectric relaxation timescales.

In the first set of 20 ensemble simulations, the plasma was initialized in a thermal Maxwellian equilibrium. To sample the force fluctuations without perturbing the bath, we introduced passive ghost particles (species \texttt{tplus} and \texttt{tminus} in the input deck) with zero current contribution (\texttt{zero\_current = T}). These passive tracers moved along ballistic trajectories at fixed velocities but did not generate their own electromagnetic fields.

We recorded the electric field fluctuations $E(t)$ experienced by these tracers. The memory kernel $\gamma(t)$ was then extracted via the inversion of the second FDT (Eq. \ref{eq:FDT}) using the Force Autocorrelation Function (FACF) averaged over the ensemble:
\begin{equation}
    \gamma(t) = \frac{Q^2}{M k_B T} \langle E(t) \cdot E(0) \rangle_{\mathrm{ensemble}}.
\end{equation}
This procedure yielded the non-Markovian friction kernel solely from equilibrium noise.

In the second set of simulations, we introduced active test particles fully coupled to the Maxwell solver, allowing them to excite wakes and experience self-consistent drag forces. To thoroughly validate the framework across different kinetic regimes, we tracked projectiles initialized at both a sub-thermal velocity ($v_0 = 0.3 v_{th}$) and a supersonic velocity ($v_0 = 3.0 v_{th}$).

\begin{figure}[htbp]
    \centering
    \includegraphics[width=\columnwidth]{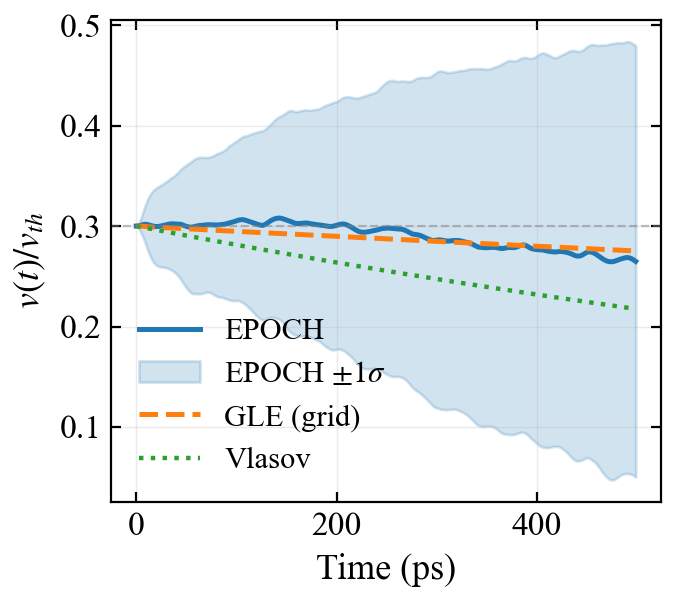}
    \caption{Comparison of the test particle velocity decay for the sub-thermal regime ($v_0 = 0.3 v_{th}$). The solid blue line represents the mean high-fidelity PIC result with a shaded region indicating one standard deviation. The dashed orange line represents the GLE prediction using the memory kernel extracted from equilibrium fluctuations. The dotted green line illustrates the theoretical Vlasov curve, which assumes a Markovian instantaneous response.}
    \label{fig:validation_subthermal}
\end{figure}

\begin{figure}[htbp]
    \centering
    \includegraphics[width=\columnwidth]{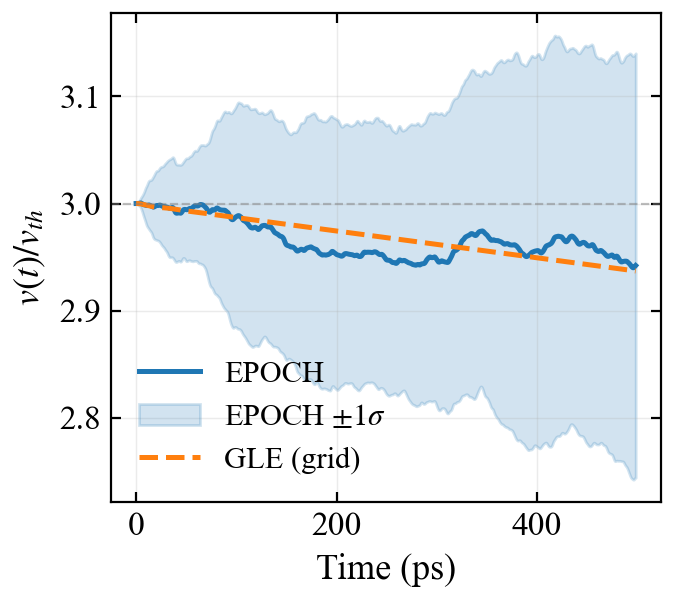}
    \caption{Comparison of the test particle velocity decay for the supersonic regime ($v_0 = 3.0 v_{th}$). The dashed orange GLE curve shows excellent agreement with the solid blue PIC simulation data, confirming the validity of the Doppler-shifted FDT at high velocities.}
    \label{fig:validation_supersonic}
\end{figure}

The validation results for the sub-thermal and supersonic cases are summarized in Fig. \ref{fig:validation_subthermal} and Fig. \ref{fig:validation_supersonic}, respectively. We compared the velocity decay observed in the brute-force drag simulations against the theoretical trajectory predicted by solving the exact GLE (Eq. \ref{eq:NonMarkovian_Accel}) using the equilibrium-extracted kernel. For the $0.3 v_{th}$ projectile shown in Fig. \ref{fig:validation_subthermal}, the GLE prediction tracks the non-equilibrium PIC results exceptionally well. Furthermore, this figure includes the Markovian Vlasov reference curve, which visibly overestimates the drag because it fails to account for the finite relaxation time of the plasma. By incorporating the full frequency-dependent structure of the memory kernel, the GLE correctly captures the transient dynamics of the plasma wake formation.

This accuracy extends to the high-speed inertial drag regime shown in Fig. \ref{fig:validation_supersonic}. The GLE successfully tracks the $3.0 v_{th}$ projectile, confirming that the equilibrium memory kernel, when properly Doppler-shifted, contains the complete nonlinear friction physics required across velocity spaces.

Beyond theoretical accuracy, the GLE framework provides a massive reduction in computational cost for modeling complex dynamic friction problems. To quantify this advantage, we compared the floating-point operations required to integrate the non-Markovian equation of motion against a standard brute-force PIC simulation using the EPOCH code. Evaluating the exact discrete convolution for the semi-implicit solver over 5000 timesteps requires approximately $5 \times 10^7$ FLOPS. In contrast, explicitly resolving the same physical time scale in a fully kinetic non-equilibrium plasma simulation necessitates tracking field updates and particle trajectories across the entire domain. For our simulation domain utilizing 768 grid cells and 307200 macroparticles, updating the field equations and advancing the particle momenta over 484000 simulation timesteps demands roughly $2 \times 10^{13}$ FLOPS. Consequently, utilizing the equilibrium memory kernel to predict stopping power reduces the required computational workload by a factor of $4 \times 10^5$. By extracting friction properties from a single thermal bath rather than running numerous high-resolution non-equilibrium simulations, this approach accelerates predictive friction calculations by over five orders of magnitude.
\section{Conclusion} \label{sec:conclusions}
In this work, we have presented a rigorous statistical mechanical framework for calculating nonlinear friction coefficients in driven systems without relying on direct non-equilibrium simulations. By reformulating the drag problem within the context of the GLE, we demonstrated that the dissipative forces acting on a high-speed projectile are not distinct dynamical entities, but are mathematically equivalent to a kinematic sampling of the inherent equilibrium fluctuations of the bath.

The central result of this study is the Doppler-Shifted FDT. We showed that for systems governed by linear response, the non-Markovian memory kernel is invariant under Galilean transformations. Consequently, the full velocity-dependent stopping power curve can be reconstructed \textit{ab initio} from the force autocorrelation function of a single stationary particle. This insight unifies the diffusive, Ohmic friction at low velocities (Stokes Law) and the inertial, wake-dominated drag at high velocities (Chandrasekhar limit) into a single geometric operation on the equilibrium spectrum.

We validated this formalism by applying it to the canonical problem of a heavy ion traversing a uniform Maxwellian plasma. We derived an explicit Gaussian memory kernel from the Vlasov dielectric response and showed that its Markovian (ballistic) limit exactly reproduces the Chandrasekhar dynamical friction formula \cite{Chandrasekhar1943}. Furthermore, we recovered the exact non-Markovian equation of motion, which captures transient history forces that are often neglected in standard kinetic models \cite{Spitzer1962, Balescu1960}. While the standard characterization of stopping power requires a brute force approach involving $\mathcal{O}(N)$ non-equilibrium simulations to map the drag curve point-by-point \cite{Zwicknagel1999}, ours reduces this cost to a single $\mathcal{O}(1)$ equilibrium simulation. Moreover, by extracting friction properties from a thermal bath rather than a violently driven wake, this method avoids common simulation artifacts such as finite-size wake wrapping and numerical heating \cite{Starrett2015}.

This framework also offers a promising path for exploring dynamic friction in regimes where analytical kinetic theories break down, such as Warm Dense Matter (WDM) and strongly coupled plasmas \cite{Murillo2000, Daligault2019}. In these systems, where the assumption of discrete binary interactions fails, the equilibrium memory kernel remains a well-defined and accessible quantity. Extending the Doppler-shifted FDT to these complex fluids could provide a first-principles method for predicting viscosities, stopping powers, and thermal relaxation rates that are critical for modern inertial confinement fusion efforts \cite{Hu2024}.
\section{Acknowledgements}
This work was supported by DARPA under Award No. N66001242007. The views, opinions, and findings expressed herein are those of the authors and do not necessarily reflect the official policy or position of the U.S. Government or any of its agencies.

\balance
\balance


\begin{thebibliography}{99}
% --- [1-6] General Non-Eq Stat Mech & FDT (The Framework) ---
\bibitem{Onsager1931} % NEW
L. Onsager, 
\textit{Reciprocal Relations in Irreversible Processes. I},
Phys. Rev. \textbf{37}, 405 (1931).

\bibitem{Green1954} % NEW
M. S. Green, 
\textit{Markoff Random Processes and the Statistical Mechanics of Time-Dependent Phenomena. II. Irreversible Processes in Fluids},
J. Chem. Phys. \textbf{22}, 398 (1954).

\bibitem{Callen1951}
H. B. Callen and T. A. Welton, 
\textit{Irreversibility and Generalized Noise},
Phys. Rev. \textbf{83}, 34 (1951).

\bibitem{Kubo1966}
R. Kubo, 
\textit{The fluctuation-dissipation theorem},
Rep. Prog. Phys. \textbf{29}, 255 (1966).

\bibitem{Evans2008}
D. J. Evans and G. P. Morriss, 
\textit{Statistical Mechanics of Nonequilibrium Liquids} 
(Cambridge University Press, Cambridge, 2008).

\bibitem{Zwanzig1961}
R. Zwanzig, 
\textit{Memory Effects in Irreversible Thermodynamics},
Phys. Rev. \textbf{124}, 983 (1961).

% --- [7-8] The Computational Bottleneck (Brute Force) ---
\bibitem{Peter1991}
Th. Peter and J. Meyer-ter-Vehn, 
\textit{Energy loss of heavy ions in dense plasma},
Phys. Rev. A \textbf{43}, 1998 (1991).

\bibitem{Zwicknagel1999}
G. Zwicknagel, C. Toepffer, and P.-G. Reinhard, 
\textit{Stopping of heavy ions in plasmas at strong coupling},
Phys. Rep. \textbf{309}, 117 (1999).

% --- [9-13] GLE Formalism & Stochastic Methods ---
\bibitem{Mori1965}
H. Mori, 
\textit{Transport, Collective Motion, and Brownian Motion},
Prog. Theor. Phys. \textbf{33}, 423 (1965).

\bibitem{Grabert1982}
H. Grabert, 
\textit{Projection Operator Techniques in Nonequilibrium Statistical Mechanics} 
(Springer, Berlin, 1982).

\bibitem{VanKampenBook} % NEW
N. G. van Kampen, 
\textit{Stochastic Processes in Physics and Chemistry} 
(North-Holland, Amsterdam, 1992).

\bibitem{GardinerBook} % NEW
C. W. Gardiner, 
\textit{Handbook of Stochastic Methods} 
(Springer, Berlin, 1985).

\bibitem{DeGrootMazur} % NEW
S. R. de Groot and P. Mazur, 
\textit{Non-Equilibrium Thermodynamics} 
(Dover, New York, 1984).

% --- [14-16] The Canonical Application (Drag Theory) ---
\bibitem{Chandrasekhar1943}
S. Chandrasekhar, 
\textit{Dynamical Friction. I. General Considerations},
Astrophys. J. \textbf{97}, 255 (1943).

\bibitem{Spitzer1962}
L. Spitzer, 
\textit{Physics of Fully Ionized Gases} 
(Interscience, New York, 1962).

\bibitem{LandauKinetics} % NEW
E. M. Lifshitz and L. P. Pitaevskii, 
\textit{Physical Kinetics (Landau and Lifshitz Course of Theoretical Physics, Vol. 10)} 
(Pergamon Press, Oxford, 1981).

% --- [17-21] Markovian vs Non-Markovian Plasma Issues ---
\bibitem{IchimaruBasic}
S. Ichimaru, 
\textit{Basic Principles of Plasma Physics} 
(Benjamin/Cummings, Reading, MA, 1973).

\bibitem{Balescu1960}
R. Balescu, 
\textit{Irreversible Processes in Ionized Gases},
Phys. Fluids \textbf{3}, 52 (1960).

\bibitem{Lenard1960} % NEW
A. Lenard, 
\textit{On Bogoliubov's Kinetic Equation for a Spatially Homogeneous Plasma},
Ann. Phys. (N.Y.) \textbf{3}, 242 (1960).

\bibitem{KlimontovichBook} % NEW
Y. L. Klimontovich, 
\textit{Kinetic Theory of Nonideal Plasmas} 
(Academic Press, New York, 1975).

\bibitem{Hahm2022}
T. S. Hahm and H. H. Kaang, 
\textit{Non-Markovian turbulent transport},
Phys. Plasmas \textbf{29}, 102302 (2022).

% --- [22-24] Connection to Modern Methods ---
\bibitem{Daligault2019}
J. Daligault, 
\textit{Calculations of the electron-ion energy relaxation rate in dense plasmas from equilibrium molecular dynamics simulations},
Phys. Rev. E \textbf{100}, 043201 (2019).

\bibitem{HansenBook}
J.-P. Hansen and I. R. McDonald, 
\textit{Theory of Simple Liquids} 
(Academic Press, London, 2013).

\bibitem{AllenTildesley} % NEW
M. P. Allen and D. J. Tildesley, 
\textit{Computer Simulation of Liquids} 
(Oxford University Press, Oxford, 1987).

% --- [25-30] Additional Context (Simulations & Viscosity) ---
\bibitem{Murillo2000}
M. S. Murillo, 
\textit{Viscosity of strongly coupled plasma},
Phys. Rev. E \textbf{62}, 4115 (2000).

\bibitem{Starrett2015}
C. E. Starrett, J. Daligault, and D. Saumon, 
\textit{Uncertainties in stopping power of dense plasmas},
Phys. Rev. E \textbf{91}, 013104 (2015).

\bibitem{Ivlev2007}
A. V. Ivlev et al., 
\textit{Non-Newtonian viscosity of complex plasmas},
Phys. Rev. Lett. \textbf{98}, 145003 (2007).

\bibitem{IchimaruStat}
S. Ichimaru, 
\textit{Statistical Plasma Physics, Vol. I: Basic Principles} 
(Westview Press, Boulder, 2004).

\bibitem{Kubo1957}
R. Kubo, 
\textit{Statistical-Mechanical Theory of Irreversible Processes. I},
J. Phys. Soc. Jpn. \textbf{12}, 570 (1957).

\bibitem{Risken}
H. Risken, 
\textit{The Fokker-Planck Equation} 
(Springer-Verlag, Berlin, 1989).

\bibitem{Arber2015}
T. D. Arber \textit{et al.},
Plasma Phys. Control. Fusion \textbf{57}, 113001 (2015).

\bibitem{Birdsall2004}
C. K. Birdsall and A. B. Langdon, 
\textit{Plasma Physics via Computer Simulation} 
(Taylor \& Francis, New York, 2004).

\bibitem{Hu2024}
S. X. Hu, V. N. Goncharov, P. B. Radha, S. P. Regan, and E. M. Campbell,
\textit{A review on charged-particle transport modeling for laser direct-drive fusion},
Phys. Plasmas \textbf{31}, 040501 (2024).

\end{thebibliography}
\end{document}